# Effects of Sol concentration on the structural and optical properties of SnO$_2$ Nanoparticle

Ali Reza Razeghizadeh[1] Iraj Kazeminezhad[2], Lila Zalaghi[1], Vahdat Rafee[1,*]

[1,*]*Department of Physics, Faculty of Science, Payame Noor University, Iran*
[2]*Department of Physics, Faculty of Science, Shahid Chamran University, Ahwaz, Iran*

Email of the corresponding author: v.rafee@gmail.com

**Abstract**
In this paper, the effects of changes in Sol concentration on the structural and optical properties of SnO$_2$ Nanoparticles are studied through the Sol-Gel method. SnO$_2$ Nanoparticles are produced from different SnO$_2$ solution concentrations (0.1, 0.3 and 0.5 mol/L) at room temperature. X-ray diffraction (XRD) and scanning electron microscopy (SEM) analysis is used to investigate the effects of changes in Sol concentration on the crystalline and surface morphology of Nanoparticles. The XRD pattern shows that the particles are in the standard tetragonal phase of SnO$_2$. The crystallites size are 11.7nm, 18.8nm, and 74.8nm, for 0.1 M, 0.3M and 0.5 M of concentration respectively. The average size of SnO$_2$ particles decreases by a reduction in the Sol concentration. The band gap value is 4.07ev, 4.28ev and 4.36ev for 0.5M, 0.3M and 0.1M of concentration respectively. The UV-visible analysis shows that decreasing the Sol concentration will cause the absorption edge shift to the shorter wavelengths, because decreasing the Sol concentration will reduce the Nanoparticles size, and smaller Nanoparticles size absorb shorter wavelengths better and also increase the band gap.

## Introduction:

Tin Oxide is an n-type wide direct band gap semiconductor about $(3.8 - 4.3 eV)$ and indirect gap about $(2.7 - 3.1 eV)$ and transparency in the visible region is about 80% [1-3] which shows that SnO$_2$ can be efficiently used as transparent conducting oxide.

The tin dioxide, which is formed under stoichiometry condition have a few charge carriers, so its resistance is high and it's expected to be a good insulator. Also, SnO$_2$ under non-stoichiometry conditions has many charge carriers, good transparency and high conductivity [4].

Research on tin dioxide is very interesting because chemical, physical and structure properties of $SnO_2$ together with optical transparency, make $SnO_2$ Nanoparticles suitable for many applications such as solid-state gas sensors [5-11], sensing and measurement advanced sensors for power generation applications [12], optical electronic devices [13,14], rechargeable Li batteries [15,16], solar cells [17,18], optical devices [19], heterojunction solar cells[20], transparent conducting electrodes [21,22], sensor array [23,24], organic light emissive devices [25], low emission glass [26] and Electrochromic glass [27].

The several interesting research works have been done on the optical properties of $SnO_2$ [28, 30] and many processes have been developed for synthesis the $SnO_2$ Nanostructures, e.g. surfactants mediate [31, 32] thermal evaporation of oxide powders [33], microwave heating method [34, 35], molten-salt synthesis, [36], carbothermal reduction [37], laser ablation synthesis [38], RF sputtering [39] and a Sol–Gel method [40, 41].

The Sol–Gel process has several advantages because of its simplicity, easily control of the film composition, safety, low cost of the apparatus and raw materials [42].

The Sol concentrations are one of the interesting factors that affect the optical properties of Nanoparticles. This is effectively done in the previous works by different methods and different oxides [43, 45].

In this paper, the researcher selects a Sol-Gel method with different concentrations of precursor Sol in the production of the $SnO_2$ Nanoparticles. After the production, the $SnO_2$ Nanoparticles, the structural properties and surface morphologies of the $SnO_2$ Nanoparticles are characterized by X-ray diffraction (XRD) and scanning electron microscopy (SEM). The researchers use the UV-visible spectrophotometer in the wavelength 190-900nm to measure the optical properties and gets the transmittance and absorption spectra.

## Experimental details

The precursor Sol of $SnO_2$ is prepared by the Sol-Gel technique as follows. The varying Sol concentration of $SnO_2$ was prepared by adding 2.302, 6.907 and 11.511 g of $SnCl_2.2H_2O$ (molar mass 225.63g/mol, Purity 98%, Merck) to 100 ml of ethanol (($C_2H_5OH$) molar mass 46.07g/mol, Purity 98%, Merck) for 0.01, 0.03 and 0.05 mol at room temperature respectively.

At first, the solution was stirred for one hour using a magnetic stirrer (Hot plate Genway 1000 models) until a clear solution is obtained without turbidity. The solution refluxed (PHYWE LEC 03.04 models) at 80 ºC for 2 hours until a transparent Sol is obtained. In order to complete the process, the transparent Sol aging for 24 hours [46,47]. Then the Sol placed in the oven at 100 ºC until the white Gel is obtained. Then the researchers continue to heat the Gel until the Gel became dry. The dry Gel is ground to obtain the soft and white powder. Finally, the powder is annealed at 500 ºC for 1 hour until $SnO_2$ Nanoparticles is obtained.

After preparing the Nanoparticles, the structural properties and surface morphologies of the $SnO_2$ Nanoparticles are characterized by X-ray diffraction (X-pert seifert3003 model by the Philips company) and scanning electron microscopy (SEM: S4160 model by the Hitachi company). The researchers use the UV-visible spectrophotometer (UV-visible: Varian-Cary5000Scan model) in the wavelength 190-900 nm to measure the optical properties.

## Results and discussion

**Microstructure of the Nanoparticle**

The structure of the $SnO_2$ Nanopowders is studied using X-ray diffraction. In order to do this, the researchers use $2\theta$ range 20–80◦ with radiation of CuKα in wavelength λ=1.5406 Å.

Fig. 2 Shows the XRD patterns of the 0.1M, 0.3M and 0.5 M $SnO_2$ Nanopowders. According to the $SnO_2$ standard spectra, the peaks have been seen in (110), (101), (211), and (301) [48]. XRD patterns show that the $SnO_2$ Nanoparticles are in the tetragonal phase. Also, it is observed that with increasing the Sol concentration, the intensity of the XRD pattern increases. The crystal size is calculated by a Scherer formula [49,50].

$$D = \frac{k\lambda}{\beta \cos\theta} \quad (1)$$

Where D is the average crystallite size, λ is the applied X-ray wavelength, and $k$=0.89 which is a constant $\theta$ is the diffraction angle in degree and $\beta$ is the full width at half maximum (FWHM) of the diffraction peak observed in radians.

Table. 1 The crystallites size of. XRD with different Sol concentration

| Sol concentration (mol/L) | crystallites size (nm) |
|---|---|
| 0.1 | 11.7 |
| 0.3 | 18.8 |
| 0.5 | 74.8 |

Observation from XRD is shown in table 1 and Fig. 2, the crystallites size of 0.1 M, 0.3M and 0.5 M are 11.7 nm, 18.8nm, and 74.8 nm, respectively. It shows that the crystallites size is increased with increasing the Sol concentration.

The surface morphology of powders, observed from SEM is shown in Fig. 3 for 0.1M, 0.3M and 0.5M of Sol concentration. The SEM is shown in Fig. 1 the average size of the grain is 18.1 nm, 33nm, and 43.2nm for 0.1M, 0.3M and 0.5M respectively. Those show that the grain size of the powders increase when the Sol concentration increases. The result of XRD and SEM are equivalent about the variation of the particle size.

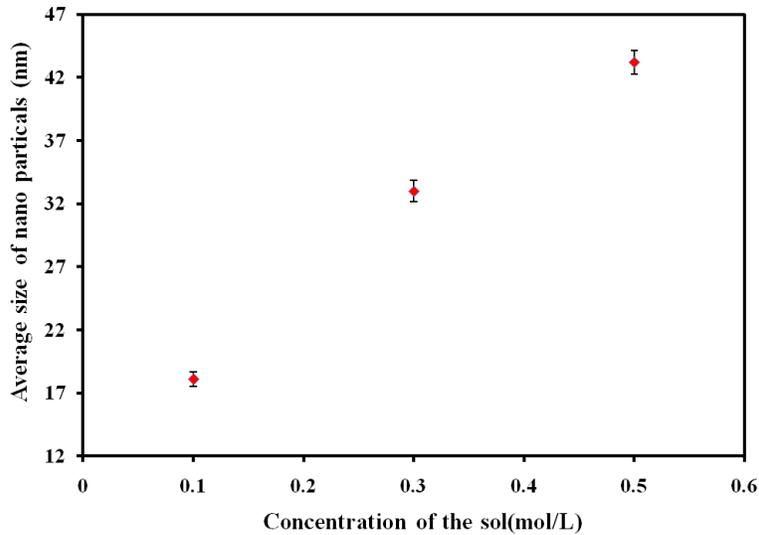

Fig.1 average size of Nanoparticles with different Sol concentration

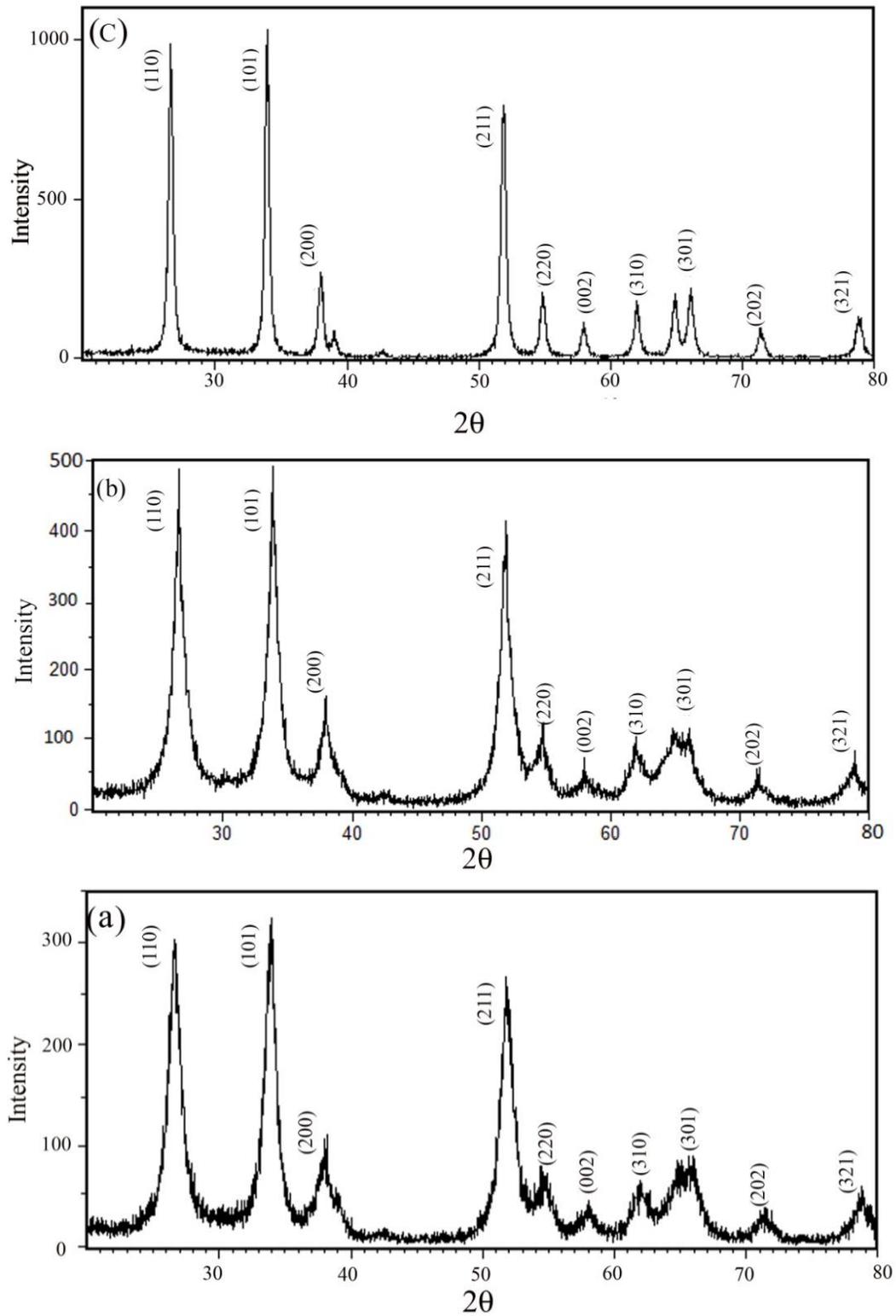

Fig. 2. Patterns of $SnO_2$ Nanoparticles with 0.1M concentration (a), 0.3M (b) and 0.5M (c).

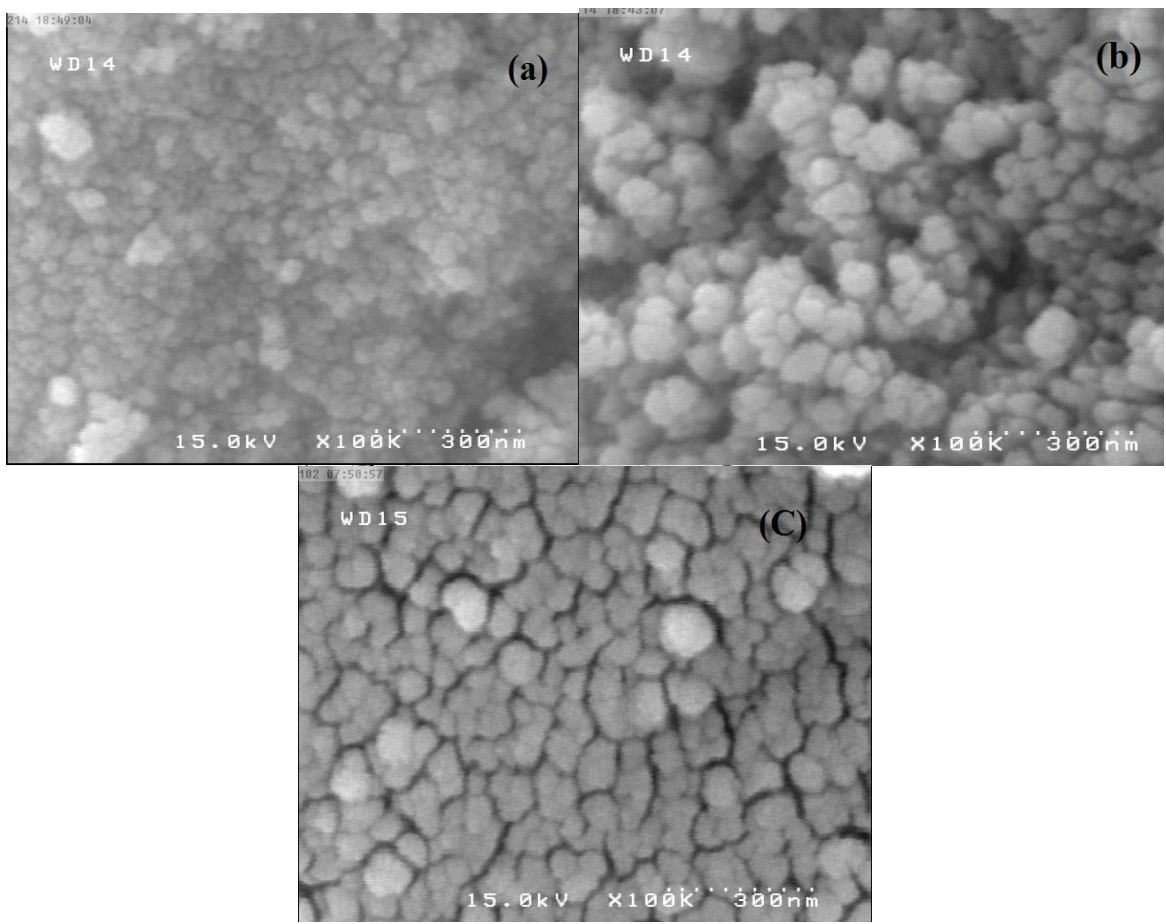

Fig. 3 SEM image of SnO$_2$ Nanoparticle fabrication with the (a) 0.1 M, (b) 0.3 M and (C) 0.5 M Sol concentration.

## Optical properties of the Nanoparticle
### FTIR Analysis

The FTIR analysis determines the functional group of materials available in the Nanopowders. The analysis is recorded using an FTIR spectrometer in a range between 380-4000 cm$^{-1}$. Fig. 4 shows the FTIR spectra of SnO$_2$ Nanoparticles with 0.1 M of the Sol concentration. The bands have been assigned due to the absorption peaks of Sn–O, Sn–O–Sn, Sn–OH and C–O, bond vibrations. In which the strong absorption band at 3419.38 cm$^{-1}$ and the band at 1633.16 cm$^{-1}$ are due to the existence of OH on the adsorbed water and Sn–OH. A sharp peak is appeared at 2341.21 cm$^{-1}$ due to carbon dioxide, which is incorporated from the atmospheric exposure. The absorption peak at 669.40 cm$^{-1}$ is assigned to Sn–O and Sn–O–Sn vibrations of SnO$_2$.

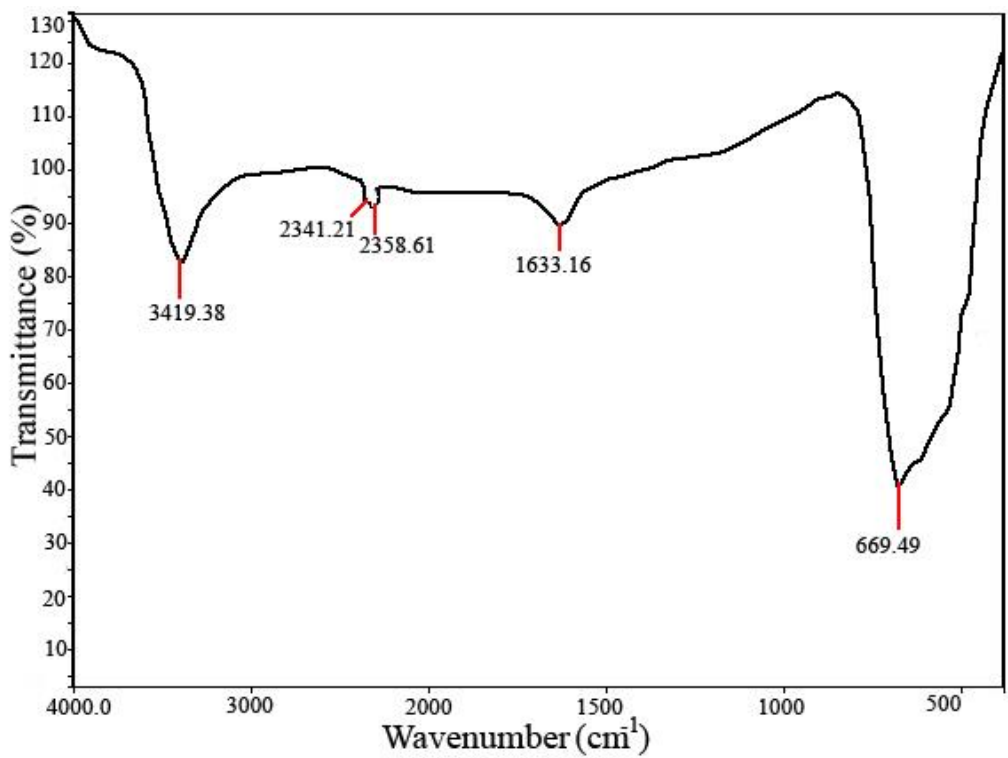

Fig. 4 The FTIR spectra of SnO$_2$ Nanoparticle with 0.1 M of Sol concentration

**UV-visible Analysis**

In order to study the optical properties of Nanoparticles, the researchers dissolved 1 g of the Nanopowders in 50 ml of distilled water. Then the solution is put in ultrasonic for 2 minutes. Then, the distilled water and the Nanopowders solution are put in absorption spectrometry. Also, the water absorption spectra of the distilled water is omitted. Finally, absorption spectra of Nanoparticles are calculated.

The researchers use the wavelength in the range of 190-900nm to dispatch photon for studying the transmittance and the absorption spectra.

The optical properties measured by UV-visible spectrophotometer (UV-visible Varian: Cary 500 scans), Fig. 5 shows the optical transmittance spectra of tin dioxide Nanoparticle with different Sol concentrations (0.1 M, 0.3 M, and 0.5 M).

SnO$_2$ Nanoparticles have high transmittance in the wider wavelength. There are three regions in the transmission curve. In the first region (250 nm $\leq \lambda \leq$ 900 nm), the transmittance increases smoothly. The second region (200 nm $\leq \lambda \leq$ 250 nm) is the transmittance falls abruptly. In the tertiary region (190 nm $\leq \lambda \leq$ 200 nm) the transmittances have an upsurge. The spectra show Nanoparticle have high transmittance in the visible region. It is observed that average transmittance in the visible region is between 65%_ 80% at different concentrations, and maximum transmittance in this region is 87.5% for 0.1M at 700 nm. It is observed that the transmittance in the visible region decreases with increasing the Sol concentration.

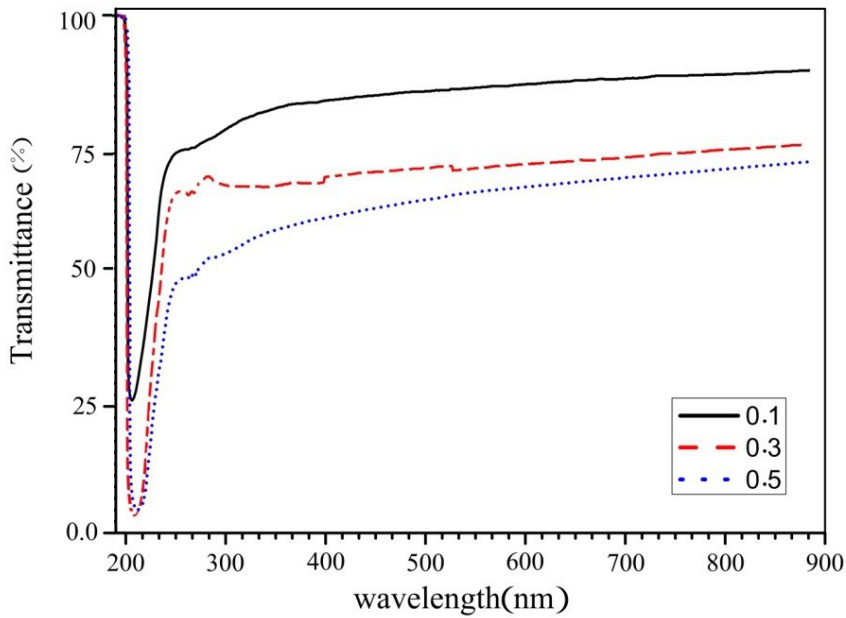

Fig. 5 the UV-visible Transmittance spectra of $SnO_2$ Nanoparticle for 0.1 M, 0.3 M and 0.5 M of concentration.

The absorption spectra are shown in Fig. 6. It shows the Nanoparticle have low absorption in the first region, have an upsurge in the second region, and in the tertiary region absorption fall abruptly. The transmittance spectrum is characterized by a sharp fall at wavelengths shorter than 250 nm, corresponding to the energy threshold for band edge absorption of $SnO_2$. The edge of absorption goes to the shorter wavelength when the concentration of the Sol decreases, because with decreasing the Sol concentration the particle size gets smaller, and the smaller particles will better absorb the shorter wavelengths.

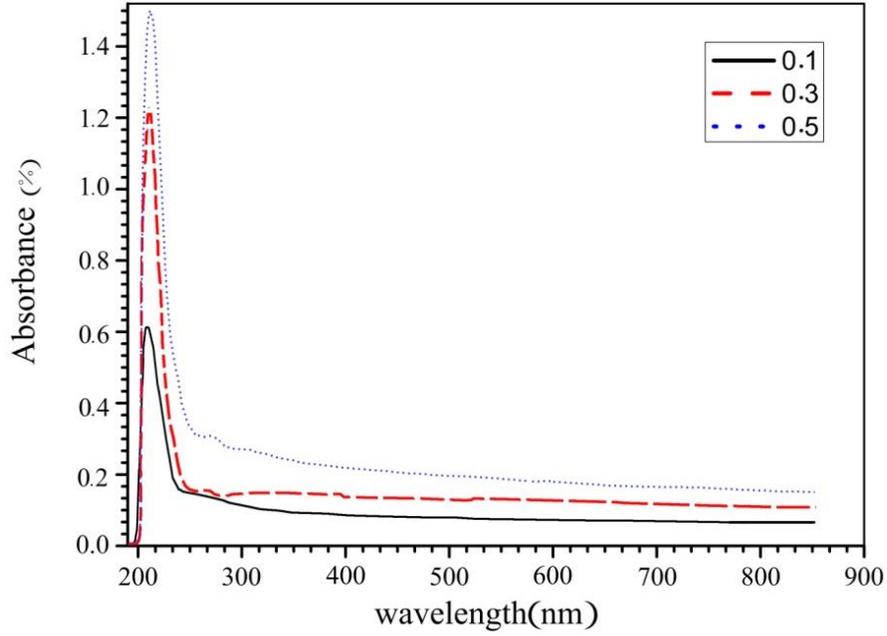

Fig. 6 the UV-visible absorption spectra of SnO$_2$ Nanoparticle with 0.1 M, 0.3 M and 0.5 M of Sol concentration.

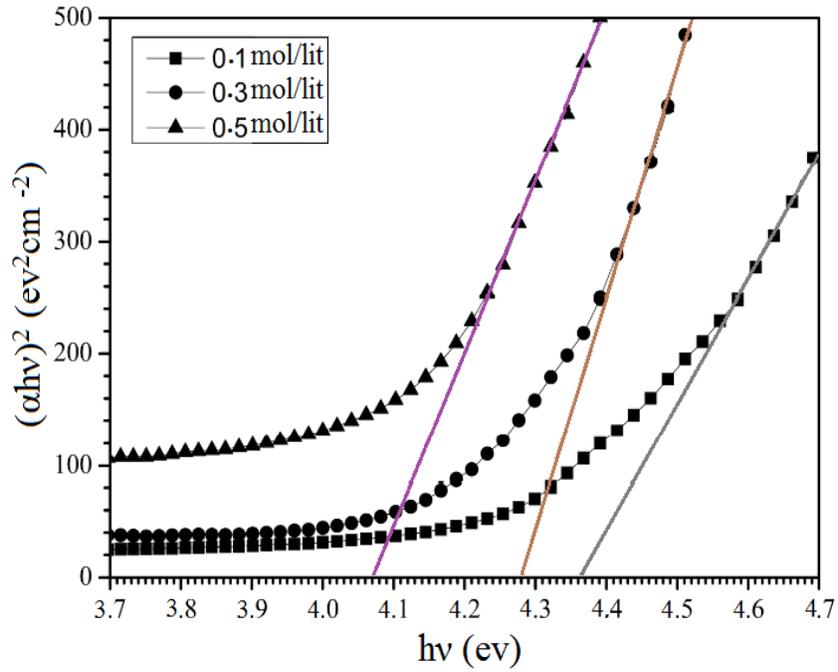

Fig. 7 Diagram of SnO$_2$ Nanoparticle band gap for different Sol concentration (0.1M, 0.3 M and 0.5 M)

From the transmittance spectra, the optical band gap energy of the Nanostructure material values can be calculated by Tauc's relation [51]:

$$(\alpha h\nu)^{\frac{1}{n}} = A(h\nu - Eg) \qquad (2)$$

$hv$ is incident photon energy, α is the absorption coefficient, *A* is a constant and *Eg* is the band gap of the material and the exponent n depends on the type of the transition. For direct and allowed transition $n=1/2$, indirect transition $n=2$ and for direct forbidden $n=3/2$ [52, 53]. For calculating the direct band gap value $(\alpha hv)^2$ versus $hv$ is plotted and it is shown in the inset of Fig. 7 by extrapolating the straight portion of the graph on $hv$ axis at α=0, so the band gap value is calculated.

The band gap value is 4.07ev, 4.28ev and 4.36ev for 0.5M, 0.3M and 0.1M of concentration respectively. The band gap value of the Nanoparticle is decreased by increasing Sol concentration. The decrease in the band gap by increasing the concentration is related to the increase of particle size.

## Conclusion

$SnO_2$ Nanoparticle prepared by the Sol– Gel method, with different Sol concentration (0.1M, 0.3M and 0.5M). XRD diffraction shows that Nanoparticles are in tetragonal phase, and the intensity of the particles increased by increasing the Sol concentration. Observations from XRD show that the crystallites size of 0.1 M, 0.3M and 0.5 M are 11.7 nm, 18.8nm, and 74.8 nm, respectively. The average size of $SnO_2$ particles decreases by reduction of the Sol concentration. The band gap value is 4.07ev for 0.5M of concentration, 4.28ev for 0.3M of concentration and 4.36ev for 0.1M of concentration. SEM images show that the average grain size is about 20-50nm, for different Sol concentration and the particle size increases when the Sol concentration increases. The optical transmittance spectra show a high transmittance in the visible region and transmittance decreases by increasing the concentration. The UV-visible analysis shows that decreasing the Sol concentration will cause the absorption edge to shift to the shorter wavelengths because decreasing the Sol concentration will reduce the Nanoparticles size and smaller Nanoparticles size absorb shorter wavelengths better and also increase the band gap.